\documentclass[preprint]{aastex62}

\usepackage[autostyle]{csquotes}
\received{}
\revised{}
\accepted{}
\shorttitle{Transient loops in the core of an active Region}
\shortauthors{Durgesh Tripathi}
\begin{document}
\title{Transient Formation of Loops in the Core of an Active Region}
\correspondingauthor{Durgesh Tripathi}
\email{durgesh@iucaa.in}
\author[0000-0003-1689-6254]{Durgesh Tripathi}
\affil{Inter-University Centre for Astronomy and Astrophysics, Post Bag - 4, Ganeshkhind, Pune 411007}
\begin{abstract}
We study the formation of transient loops in the core of the AR 11890. For this purpose, we have used the observations recorded by the Atmospheric Imaging Assembly (AIA) and the Interface Region Imaging Spectrograph (IRIS). For photospheric field configuration, we have used the line-of-sight (LOS) magnetograms obtained from the Helioseismic and Magnetic Imager (HMI). The transient is simultaneously observed in all the UV and EUV channels of AIA and the three slit-jaw images from IRIS. The co-existence of the transient in all AIA and IRIS SJI channels suggests the transient's multi-thermal nature. The transient consists of short loops located at the base of the transient as well as longe loops. A differential emission measure (DEM) analysis shows that the transient has a clumpy structure. The highest emission observed at the base is within the temperature bin of $\log\, T = 6.65 - 6.95$. We observe the longer loops at a similar temperature, albeit very feeble. Using LOS magnetograms, we conclude that the magnetic reconnection may have caused the transient. Our observations further suggest that the physics of the formation of such transients may be similar to those of typical coronal jets, albeit in different topological configurations. Such multi-wavelength observations shed light on the formation of hot plasma in the solar corona and provide further essential constraints on modeling the thermodynamics of such transients.
\end{abstract}

\section{Introduction} \label{sec:intro}
The existence of multi-million degree hot plasma in the solar corona has puzzled astronomers since its discovery in the 1940s. In the last several decades, there has been a great advancement in observational techniques and computation resources, which has led to tremendous progress in understanding this issue. However, the final solution remains elusive \citep[][]{Klim_2006, Rea_2014}. In general terms, the on-disk corona is covered by three different areas, viz. active regions, quiet Sun, and coronal holes. Of which, active regions have higher temperature followed by the quiet Sun and then coronal holes \citep[see, e.g.,][]{DelM_2018}.

Active regions show pronounced heating in the observations taken so far with imaging and spectroscopic instruments. Therefore, they have been used as an excellent target of opportunity to address solar coronal heating. The high resolution observation show that active regions are comprised of loops observed at different length scales \citep[see e.g.][and references/citations therein]{War_2008,TriMDGY_2009, Del_2013, Rea_2014,GhoT_2017}: such as fan loops (formed at $<$1~MK), warm loops (formed at $\sim$1~MK) and hot core loops ($\sim$3{--}5~MK). In addition, there is significant significant amount of diffuse emission \citep[]{DelM_2003, ViaK_2011}, which has a peak emission at $\log\,T = 6.2$ \citep[][]{SubTKM_2014, Bro_2019}. Observations of such loop structures along with the diffuse emission add further complication for the modeling.

With modern spectroscopic instruments observing in the spectral bands, which are sensitive to solar coronal plasma, the fan loops and warm loop were resolved \citep[][and references therein]{War_2008, TriMDGY_2009, GuaRP_2010, GhoT_2017}. Therefore, it has been possible to directly measure plasma parameters such as electron density, temperature, filling factors in those loops and compare with hydrodynamic modelling \citep[see e.g.][]{DelM_2003,War_2008,TriMDGY_2009,GupTM_2015,GhoT_2017}. Based on the observations from spectroscopic and imaging instruments, it has been shown to a great extent that the properties of warm and fan loops may be explained by low-frequency impulsive heating occurring in the corona.

However, in the core of the active region, structures are not well resolved, and therefore such analysis is not possible. Thus, the formation of hot plasma in the active region's core remains a highly debatable topic. Due to the unresolved nature, the authors have relied on the measurements such as Doppler shifts as well as Differential Emission Measures (DEM) \citep[see, e.g.,][]{TriMK_2010, TriKM_2011, WarWB_2012, DelTM_2015} that did not require the structures to be resolved. 

With the high-resolution observations recorded by IRIS and AIA, it has now been possible to detect individual transient events, including in the core of active regions \citep[see, e.g.][]{TesDA_2014, TesDH_2016, ChiPS_2017, ChiPS_2018}. A detailed study of the formation and thermodynamics evolution of such transients will help us comprehend the existence of the hot plasma in the core of active regions.

\cite{TesDA_2014, TesDH_2016} suggested that these transients in the core of active regions result from coronal nanoflares with non-thermal electron beams. However, \cite{ChiPS_2017, ChiPS_2018, ChiPPS_2020} suggest that these events are due to the magnetic reconnection at the footpoints driven by photospheric magnetic flux cancellation occurring within the mixed polarity regions \citep[][]{PriCS,SynPC}. However, these studies did not focus on the thermodynamic evolution of these transients \citep[][]{GupST_2018}.

In this paper, we study the transient formation of hot loops in the core of an active region. For this purpose, we have used the observations recorded with Interface Region Imaging Spectrometer \citep[IRIS;][]{iris}, Atmospheric Imaging Assembly \citep[AIA;][]{aia} and Helioseismic and Magnetic Imager \citep[HMI;][]{hmi}. We perform the full thermodynamic study of the transient loops by obtaining the differential emission measure. We also study the possible cause of the transient occurrence by using the line-of-sight (LOS) magnetograms. The rest of the paper is structures as follows. In \S~\ref{sec:obs} we describe the observations. In \S~\ref{sec:res} we describe data analysis and results. We provide a summary of the results and conclude in \S~\ref{sec:con}.

\section{Observations and data} \label{sec:obs}

For this work, we have primarily used the observations recorded by IRIS and AIA. We have also used the line-of-sight (LOS) magnetograms obtained from HMI to infer the underlying magnetic field structure. Both AIA and HMI are telescopes onboard the Solar Dynamics Observatory (SDO). All three instruments simultaneously observed the active region \textsl{AR~11890} and detected the transient formation of a loop system within the core of the active region. We note that the transient loop system studied here resembles the structure similar to those described by \cite{Han_1997} using the observations recorded by the Soft X-ray Telescope \citep[SXT;][]{TsuAB_1991} onboard \textsl{Yohkoh}. These were referred to as \enquote{double loop configuration of solar flares}. We emphasize that \cite{Han_1997}, however, only studied the morphology of these structures and did not perform the study of their thermodynamical evolution.

\begin{figure*} 
\centering
\includegraphics[width=0.8\textwidth]{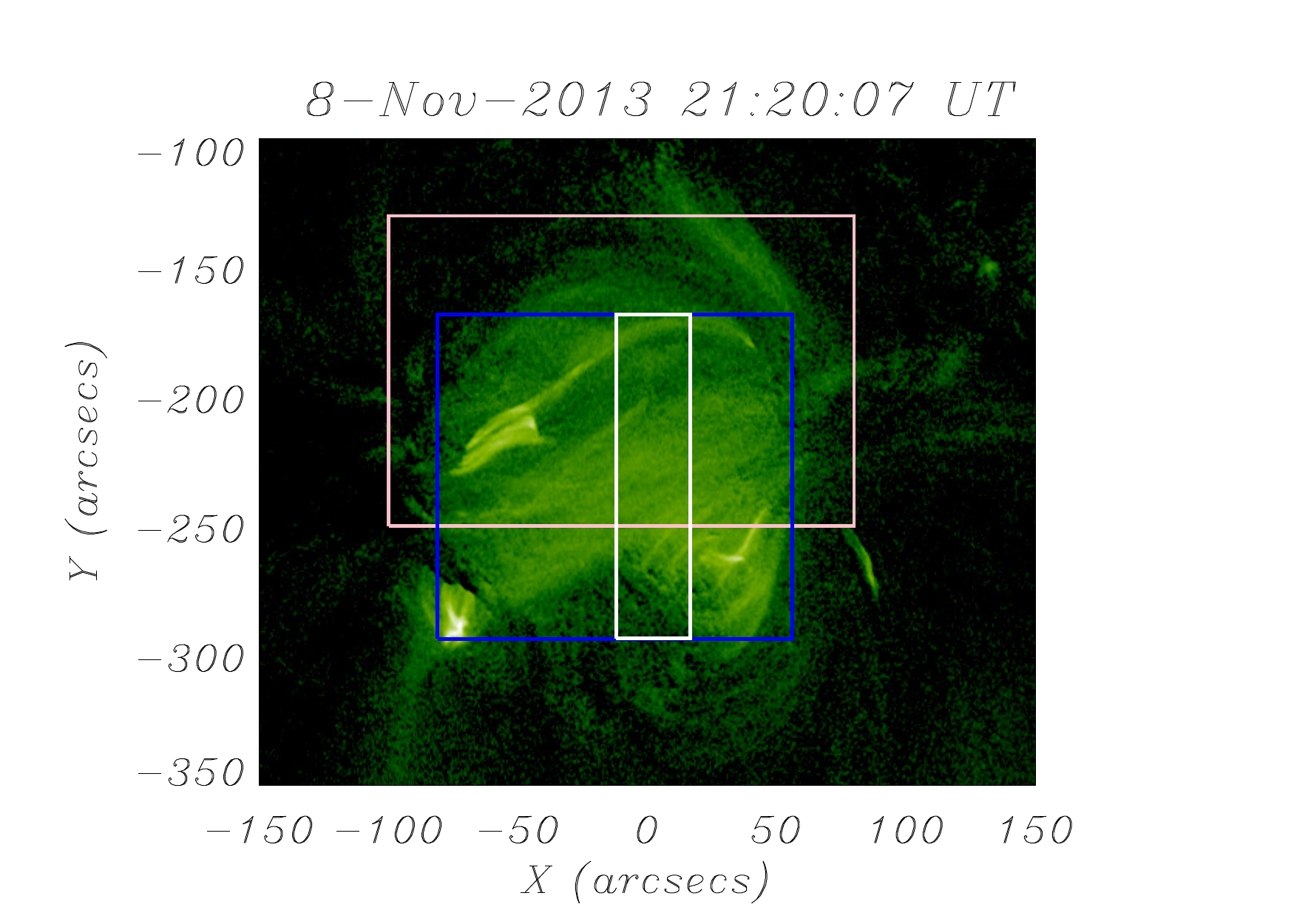}
\caption{The transient at its peak phase as observed in \ion{Fe}{18} line deduced from AIA 94~{\AA}. The pink over-plotted box represents the AIA field-of-view (FOV) that is considered for further analysis. The blue and white boxes locate the IRIS SJI and spectroscopic FOV.}\label{context}
\end{figure*}

The background image shown in Fig.~\ref{context} displays a portion of the Sun's disk covering the active region recorded at the peak of the transient loops as observed in AIA~94~{\AA} \footnote{94~{\AA} image shown in the background is obtained after removing the contribution of cooler plasma.}. The over-plotted pink box covers the region we have selected from the full disk AIA observations for further study. The area enclosed with the blue and white boxes locates the IRIS field-of-view (FOV) in SJI and spectroscopic mode. As is conspicuous, in SJI mode, IRIS covered the full extent of the feature. Unfortunately, however, the transient loops were not observed in the spectroscopic mode. However, there is still important information that warrants a detailed study using AIA and IRIS-SJI. 

In SJI modes, IRIS observes the Sun in one near UV (NUV; 2796~{\AA}) and two far-UV (FUV; 1330~{\AA}, and 1400~{\AA}) channels with a spatial resolution of 0.33~{\arcsec}. These three filters cover the solar chromosphere and transition regions, respectively. For more details of the IRIS and its capabilities, see \cite{iris}. 

The AIA, being the full disk imager, observed this event in all its filters with a pixel size of $\sim$0.6~{\arcsec}. In the current study, for the purpose of dynamic evolution, we have used the observations recorded in 1600~{\AA}, 304~{\AA}, 171~{\AA}, 193~{\AA}, 211~{\AA}, 335~{\AA}, 94~{\AA} and 131~{\AA}. Together, these filters cover a temperature range of $\log\,T = 4.8$ to $\log\,T=20$~MK. For more details on AIA and its capabilities see \cite{aia} and information on the sensitivity of different passbands to temperature \citep[][]{aia_res1,aia,aia3}. 

As demonstrated by \cite{aia_res1}, the different AIA channels have contributions from plasma at different temperatures. The three channels, namely 94~{\AA}, 335~{\AA} and 131~{\AA}, which are designed with a primary aim of studying the evolution of hot plasma, suffer from the fact that they have contributions from plasmas at lower temperatures. A couple of empirical relations, giving similar results, have been suggested \citep[see][]{cool_remove1,cool_remove2} to estimate and remove the contribution of lower temperature plasma. In the present work, we have used the method proposed by \cite{cool_remove1} to remove the cooler temperature emission from 94~{\AA} channel and studied the evolution of the transient in the \ion{Fe}{18} line.

To study the thermodynamic evolution of plasma within different temperature bins simultaneously, we have employed \textit{Differential Emission Measure (DEM)} technique developed by \cite{mark_dem}. The method developed by \cite{mark_dem} is considerably faster than the other existing codes \citep[see also][]{han_dem}. Hence, this technique allows us to study the thermodynamic evolution of the plasma in transient events. To derive the emission measure, we have used the original level~1 AIA observations for all the six EUV channels and applied aia\_prep.pro.

\section{Data Analysis and Results} \label{sec:res}
\subsection{Formation and evolution of the transient loops} \label{cool_hot}

\begin{figure*} 
\centering
\includegraphics[width=0.9\textwidth]{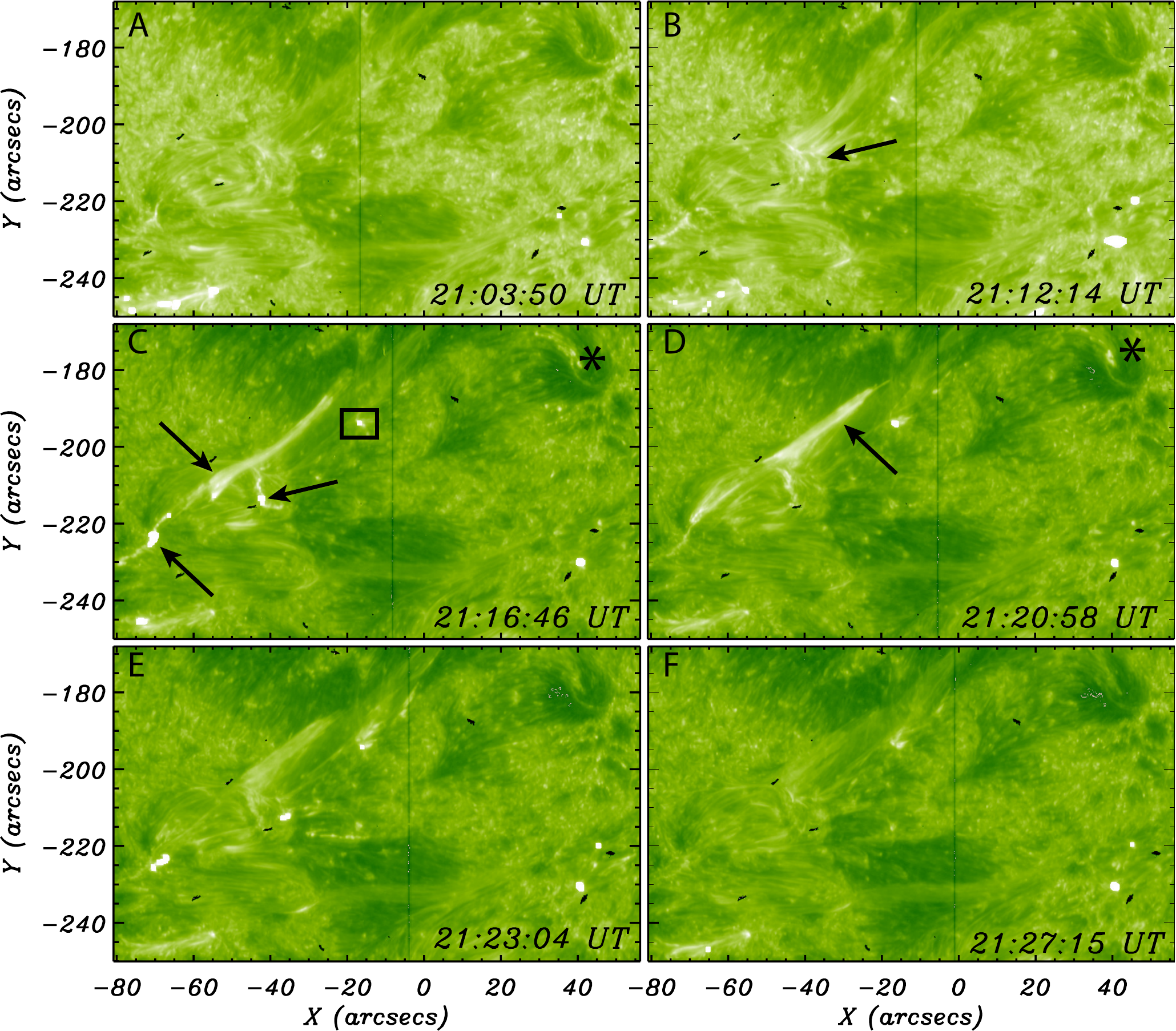}
\caption{Evolution of the transient as observed in IRIS 1400~{\AA} SJI. The arrows in panels B, C and D marks the first sign of activity, the three intense brightenings and the transient structure, respectively. The box in panel C locates another brightening that also occurs simultaneous to those marked by arrows.}\label{si4_evol}
\end{figure*}

In Fig.~\ref{si4_evol}, we display a sequence of IRIS \ion{Si}{4} images, showing the full evolutionary sequence (i.e., the formation, evolution, and eventual disappearance) within a duration of $\sim$~15 minutes of the transient loops in the core of the active region. We observe the first brightening at the location marked with the arrow shown in Fig.~\ref{si4_evol}.B. In the next frame, recorded after about 4 minutes, we observe two localized brightenings along with an inverted and slanted Y-shaped structure. This structure connects the two localized brightenings and extends further towards the west. Simultaneously, we observe another localized brightening towards further west, enclosed by a box in Fig.~\ref{si4_evol}.C, which evolves in subsequent images. The transient structure attains its maximum brightness after about 8 minutes from the onset (see Fig.~\ref{si4_evol}.D) and disappears completely in panel E, with the footpoint localized brightening still prevalent. At the peak of the transient, another simultaneous brightening is observed at [40, -180], shown by asterisks in Figs.~\ref{si4_evol}.C and D.

To study the cooler counterpart of the transient, in Fig.~\ref{cool} we display the observations recorded by IRIS in \ion{Mg}{2} \& \ion{C}{2} (panels A \& C) and those recorded by AIA in 304~{\AA} \& 1600~{\AA} (panels B \& D). These images are taken at the peak of the transient. Note the larger FOV of AIA images compared to that of IRIS images. We selected the bigger FOV of AIA images to obtain a more comprehensive view of the transient. The inverted and slanted Y-shaped structure of the transient, including the footpoint brightening, is well observed in all the channels. Moreover, associated with the transient, the 304~{\AA} channel image reveals the presence of a twisted filament. The filament is not observed in \ion{Mg}{2} and \ion{C}{2} due to smaller FOV, albeit the western footpoint. The isolated localised brightening enclosed by the box in Fig.~\ref{si4_evol}.B) is also seen in \ion{Mg}{2} as well as \ion{C}{2}. However, no such brightening is observed in 304~{\AA} and 1700~{\AA} images. The brightening located with asterisks at [40, -180] in Figs.~\ref{si4_evol}.C and D is also conspicuously seen in Fig.~\ref{cool}.C and D. We note that this brightening is located at the western footpoint of the associated filament (marked with an arrow). The 304~{\AA} image and the corresponding animation (Figure3D\_animation.mp4) demonstrate that the transient started at the eastern footpoint of the filament. We observe mass motion, i.e., movement of brightenings in both directions along the filament axis. We further note the brightenings of some strands winding around the filament. 

The corresponding coronal images recorded at the peak of the evolution are shown in Fig.~\ref{hot}. We note that the transient is well observed in all the AIA channels, particularly the transient base. The filament structure observed in 304 image is also visible in 171~{\AA}, 193~{\AA}, 211~{\AA} as well as 131~{\AA} (mostly due to the contribution of \ion{Fe}{8} line). The 335 and 94~{\AA} images suggest that the base of the transient comprises multiple loops; only some are seen in lower temperature filters. The 94~{\AA} images show the transient in its complete virtue. The base of the transient comprises multiple loops that are crisscrossing and a long loop that is connecting the base of the transient to the western footpoint of the filament.

To get a better understanding of the hot loops in the transient, we have employed the method of \cite{cool_remove1} on 94~{\AA} images taken by AIA to obtain \ion{Fe}{18} emission. Fig.~\ref{hot_94} and corresponding animation (Figure5\_animation.mp4) displays a sequence of thus obtained images representing \ion{Fe}{18} emission with peak formation temperature of $\log\, T [K]$ = 6.85. Note that we have displayed these images in negative intensities. Fig.~\ref{hot_94} shows that the transient base consists of loops connecting the two footpoints (see also the corresponding animation). The two footpoints brighten up nearly simultaneously (see frame number 15 in Figure5\_animation.mp4).

With passing time, a few more loops appear. In frame number 40 onwards, we identify a long loop-like structure (see Fig.~\ref{hot_94}.A) that connects the western foot point. The loop evolves with time, and the width increases, suggesting it to be a multi-stranded structure. We also note a kink at [-40,-200], located with an arrow in Fig.~\ref{hot_94}.E. The presence of the kink suggests that there are loops connecting the kink location with the transient western footpoint and the base. Finally, the complete structure fades aways. The total time of evolution of the transient is 15~mins. The images displayed in Fig.~\ref{hot_94} also demonstrate that the transient is multi-stranded and different strands are lighting up at different times in the course of the evolution.

\subsection{DEM inversion and temperature evolution}\label{dem}

To study the temperature structure of the transient, we have performed emission measure (EM) analysis using the six coronal channels of AIA. For this purpose, we have employed the EM code developed by \cite{mark_dem}. The EM is computed for each pixel. Note that to derive the EM, we have used original level 1.5 data from all the channels including 94~{\AA}.

In Fig.~\ref{em_maps} (see also the corresponding animation Figure6\_animation.mp4), we plot the EM-maps obtained in different temperature bins as labeled at the peak of the transient (corresponding to the AIA images shown in Fig.~\ref{hot}). The EM maps demonstrate the clumpy nature of the transient, both at the base as well as along the loop-like structure connecting the western footpoint. At the peak phase, the base of the transient is brightest in the EM map obtained for $\log\,T=[6.35-6.95]$ (Figs.~\ref{em_maps}.C \& D). However, in this temperature range, there is almost no emission from the western side of the transient. The EM maps further suggest that the base and lower half of the transient are multi-thermal in nature. We further note that the base, as well as the long loop-like structure, are thinner at higher temperatures, suggesting that only a few strands are at higher temperatures. The long loop structure is clearly and completely seen in the EM-maps corresponding to temperature bins of $\log\,T=[6.95-7.25]$ and $\log\, T=[7.25- 7.55]$ (see Figs.~\ref{em_maps}.E \& F). This EM maps suggest that the long loop structure also seen in \ion{Fe}{18} images (see Fig.~\ref{hot_94}.D) might be as high as $\log\, T = [7.25 - 7.55]$. However, we note that the emission measure for the long loop in this temperature range is very low. 

To obtain the quantitative temperature structure of the transient, we plot the EM curves obtained at seven different locations [P0, P1... P6], as shown in Fig.~\ref{em_plots}. The plots are obtained at three different instances during the evolution of the transient. The blue curves show the EM during the early phase, whereas the orange and red curves are at the peak and towards the end of the transient. The plots show that at the peak of the transient, there is a significant enhancement in EM. The EM curves for P0, P1, and P2 peaks at around $\log\, T=6.5$, which is usually seen for the core of active regions \citep[see, e.g.][]{TriKM_2011, Del_2013}. For the rest of the locations, at the maximum phase of the transient, the peaks of the EM curves shift towards higher temperatures. For locations P3 and P4, the EM curves have long tails towards the higher temperature, suggesting that it contains plasma at much higher temperatures.

It is important to note that although the long loop structure is observed, albeit very faint, in the EM map corresponding to the temperature bin of $\log\, T = 7.25 - 7.55$ (Fig.~\ref{em_maps}.F), it is not reflected in the EM curves shown in Fig.~\ref{em_plots}. This could be because of not well constrained EM curves towards the higher temperatures \citep[see, e.g.,][]{WinWS_2012}. It is also important to emphasize that the EM maps obtained at high temperatures, in particular in the bins of $\log\, T = 6.95 - 7.25$ and $\log\, T = 6.25 - 7.55$, are very likely erroneous, as was suggested by \cite{YouM_2014} and may be due unaccounted cooler lines present within the passbands of 94~{\AA} and 131~{\AA}. Therefore, caution must be exercised while interpreting such EM maps obtained at higher temperatures.

\subsection{Evolution of the Photospheric Magnetic field at the base of the transient}\label{mag}

In order to study the origin of the transient, we combine the AIA observations with the LOS of magnetograms recorded with HMI. Fig.~\ref{hmi_aia}.A shows an HMI LOS magnetogram in the background and AIA 94~{\AA} image in the foreground, both taken nearly simultaneously at the peak of the transient. In Fig.~\ref{hmi_aia}.B, we have over-plotted contours ($\pm$100~G) of magnetic flux density on AIA 94~{\AA} image. The figure clearly shows that the two footpoints of the transient base are rooted in two different polarity regions. The long loop-like structure connects to the other major polarity of the active region.

To study the time evolution of the source region magnetic field, we have also plotted a sequence of HMI LOS magnetograms in Fig.~\ref{hmi_mag}. These are the closest in time for the AIA~94{\AA} images shown in Fig.~\ref{hot_94}. The two footpoints of the transient base are located by two arrows in the top left panel. As can be readily seen, the base location has an abundance of mixed polarities. The first brightenings, which are also the footpoints of the base of the transient, seen in \ion{Si}{4} images shown in Fig.~\ref{si4_evol} occur at these locations. Therefore, it is plausible to conclude that those brightenings and the eventual transient occurred due to heating caused by the magnetic reconnection. However, we note that in 19 mins for which LOS magnetograms are displayed, we do not observe any significant change. 

We have also studied the longer-term evolution of magnetic flux. For this purpose, we have obtained HMI magnetograms starting from 18:00 UT (almost three hours before the trigger of the transient). These images are shown in Figure9\_animation.mp4, which demonstrates changes in the magnetic field configuration in the footpoint region. To quantify any change, if at all, in the photospheric magnetic flux at the footpoint regions, we study the evolution of magnetic flux starting three hours before the launch of the transient. For this purpose, we chose a small region located at the right footpoint of the base, as shown in the bottom right panel of Fig.~\ref{hmi_mag}. In Fig.~\ref{mag_flux}, we plot the evolution of total signed flux (blue), total positive flux (orange), and total negative flux (red). We note that for the first one hour, both the positive flux and negative flux decreased, which is also reflected in the decrease in the total signed flux. Beyond 19:00 UT, positive flux remains almost constant, and negative flux continues to decrease until the transient trigger, also reflected in the total signed flux. Therefore, it is plausible to conclude that such a change in the field configuration may have been central to the trigger of the transient. However, it is important to note that the box over which the flux is computed may not be representative. Moreover, the question remains: should it really take three hours of continuous change in the photospheric flux to create such transients in the corona.

\section{Summary and Conclusions} \label{sec:con}

In this paper, we present a multi-wavelength study of the formation of a transient event in the core of an active region. For this purpose, we have used slit-jaw images from IRIS and all the EUV and 1600~{\AA} images of AIA. To probe the transient's thermodynamic nature, we have employed the DEM technique on the EUV observations of AIA taken in 6 coronal channels. To shed light on the transient's possible origin, we have also studied the LOS magnetograms obtained by HMI.

The transient is observed in the slit-jaw images of \ion{Mg}{2}, \ion{C}{2} and \ion{Si}{4} recorded using IRIS as well as in the UV and EUV channels of AIA. The two footpoints of the transient brightens-up nearly simultaneously, as is noted from SJI's recorded in \ion{Si}{4}. Such simultaneous brightenings at both the footpoints have also been observed by, e.g., \cite{TesDA_2014, ChiPS_2017, GupST_2018}. 

\begin{figure*}
\includegraphics[width=1.0\textwidth]{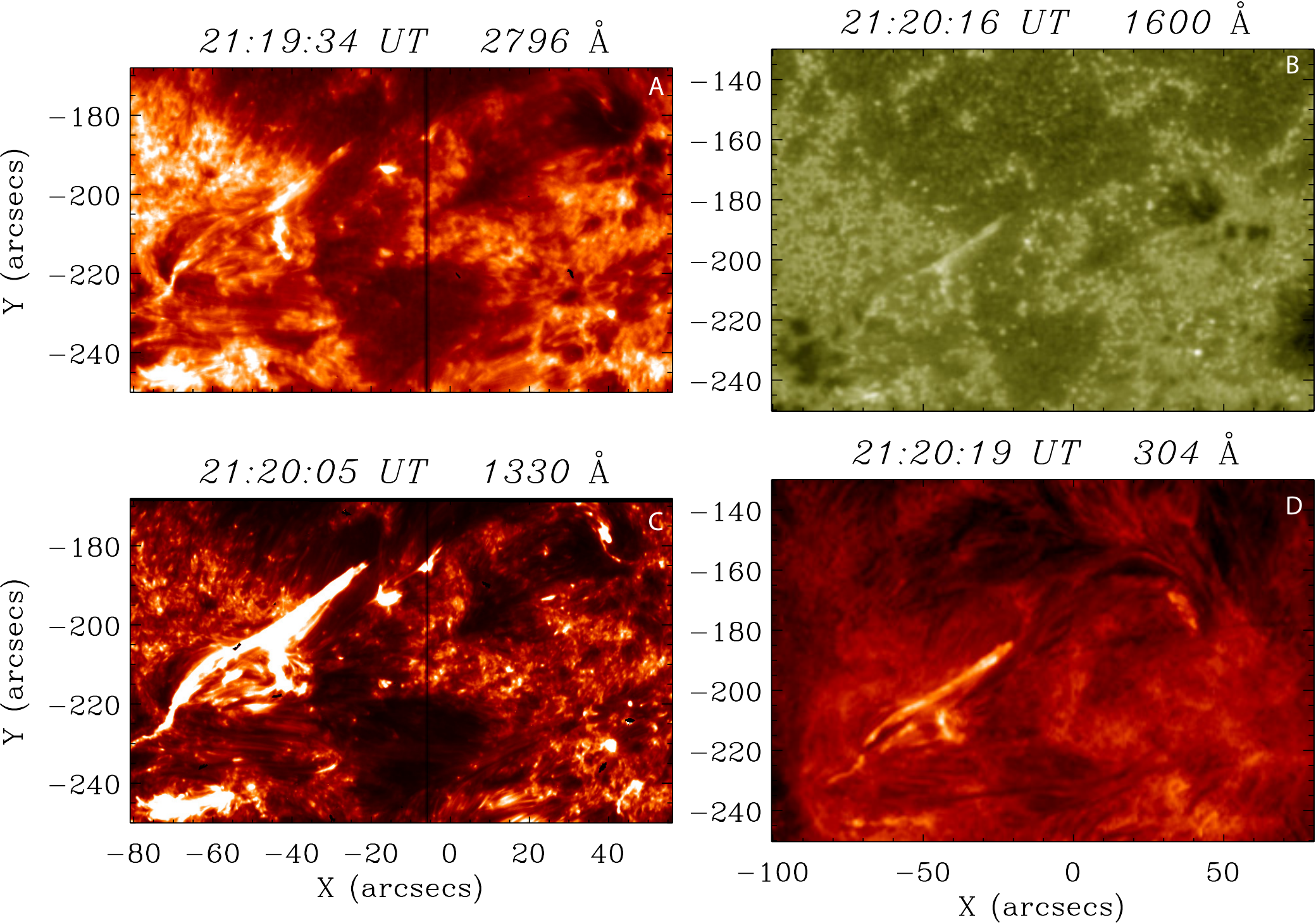}
\caption{The transient at the peak of its evolution as observed with IRIS \ion{Mg}{2} and \ion{C}{2} SJI (panels A and C) and AIA 1600 and 304~{\AA} passbands (panels C and D). An animated version of panel D is available in the HTML version of the article, showing the evolution of the transient for about 18 mins as observed in 304~{\AA} and the presence of an associated filament. Mass motions in both directions along the filament axis are seen. The animation also shows the brightenings of some strands winding around the filament.}.\label{cool}
\end{figure*}

\begin{figure*}
\includegraphics[width=1.0\textwidth]{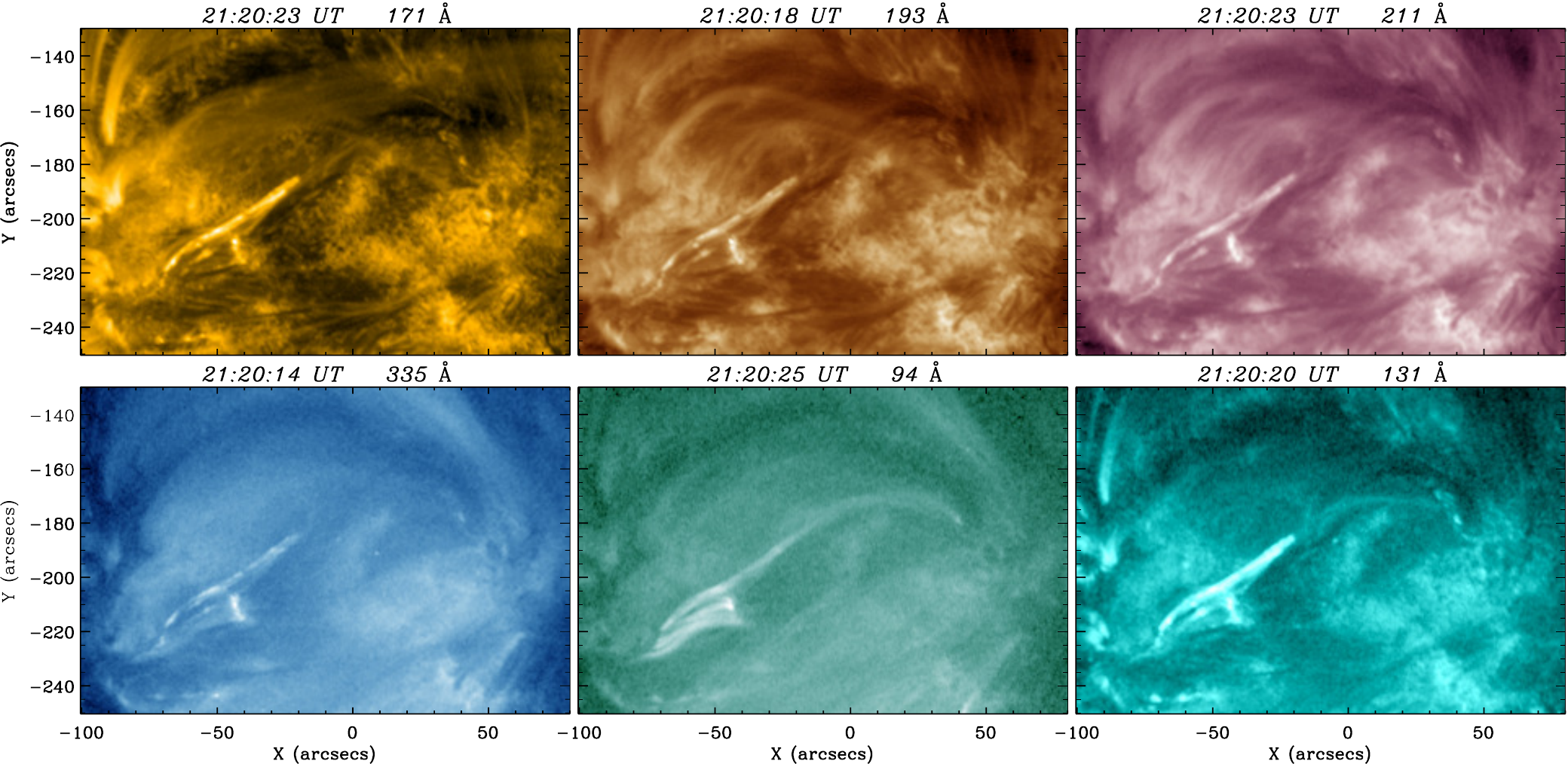}
\caption{Same as Fig.~\ref{cool} but for coronal channels of AIA as labelled.} \label{hot}
\end{figure*}

\begin{figure*}
  \centering 
  \includegraphics[width=0.95\textwidth]{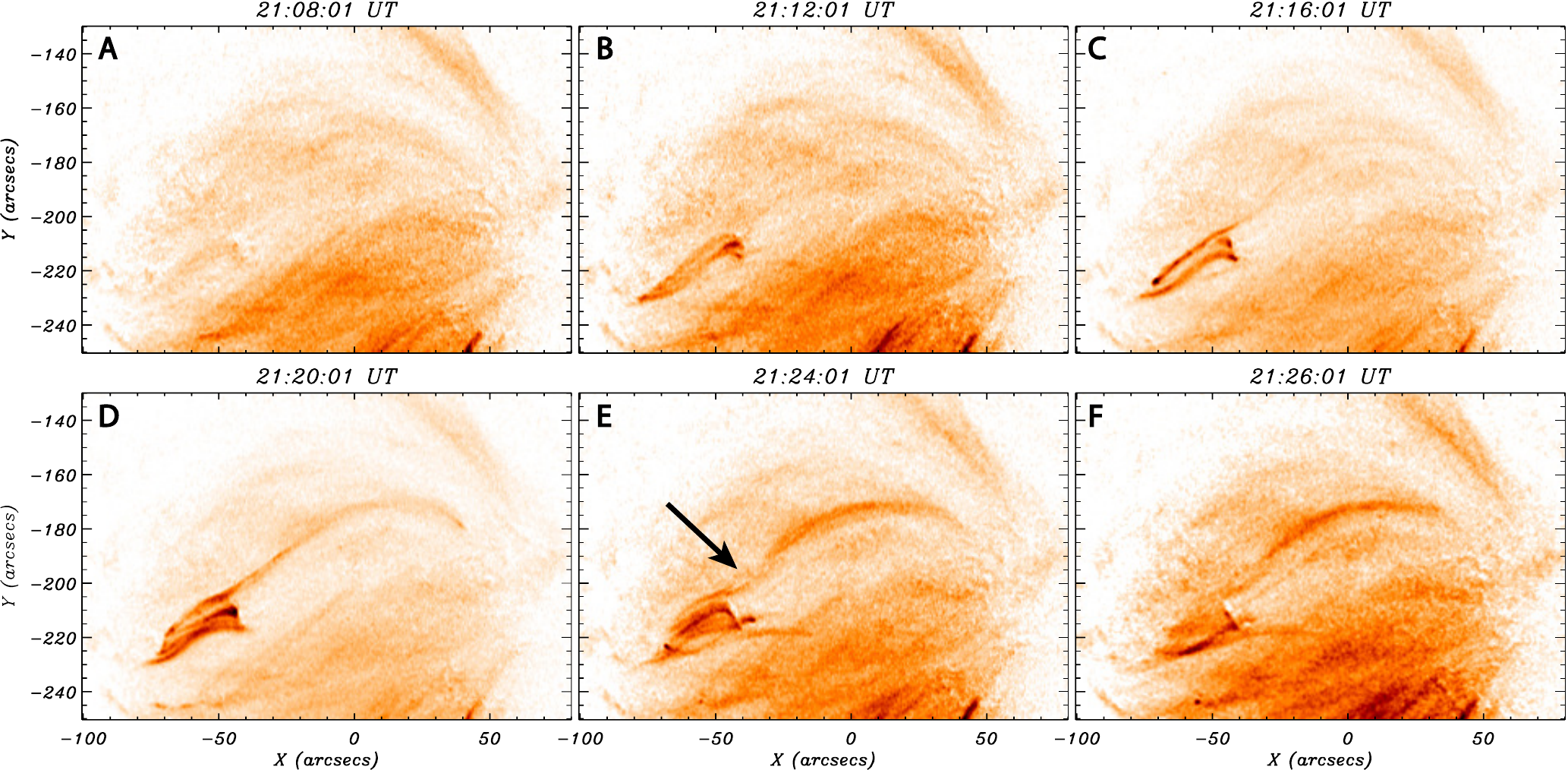}
  \caption{Evolution of the transient as observed in \ion{Fe}{18} line deduced from 94~{\AA} channel of AIA. An animated version of this figure is available in the HTML version of the article, which demonstrates the evolution of the transient for 18~min as observed in \ion{Fe}{18} line deduced from 94~{\AA}. The animations demonstrate the multi-stranded nature of the transient, with loops at different length scales. During the early phase of the evolution, loops at the base are seen. However, after about 8~mins, a longer loop structure connecting to the western footpoint is observed.}\label{hot_94}
\end{figure*}

\begin{figure*}
  \centering 
  \includegraphics[width=0.8\textwidth]{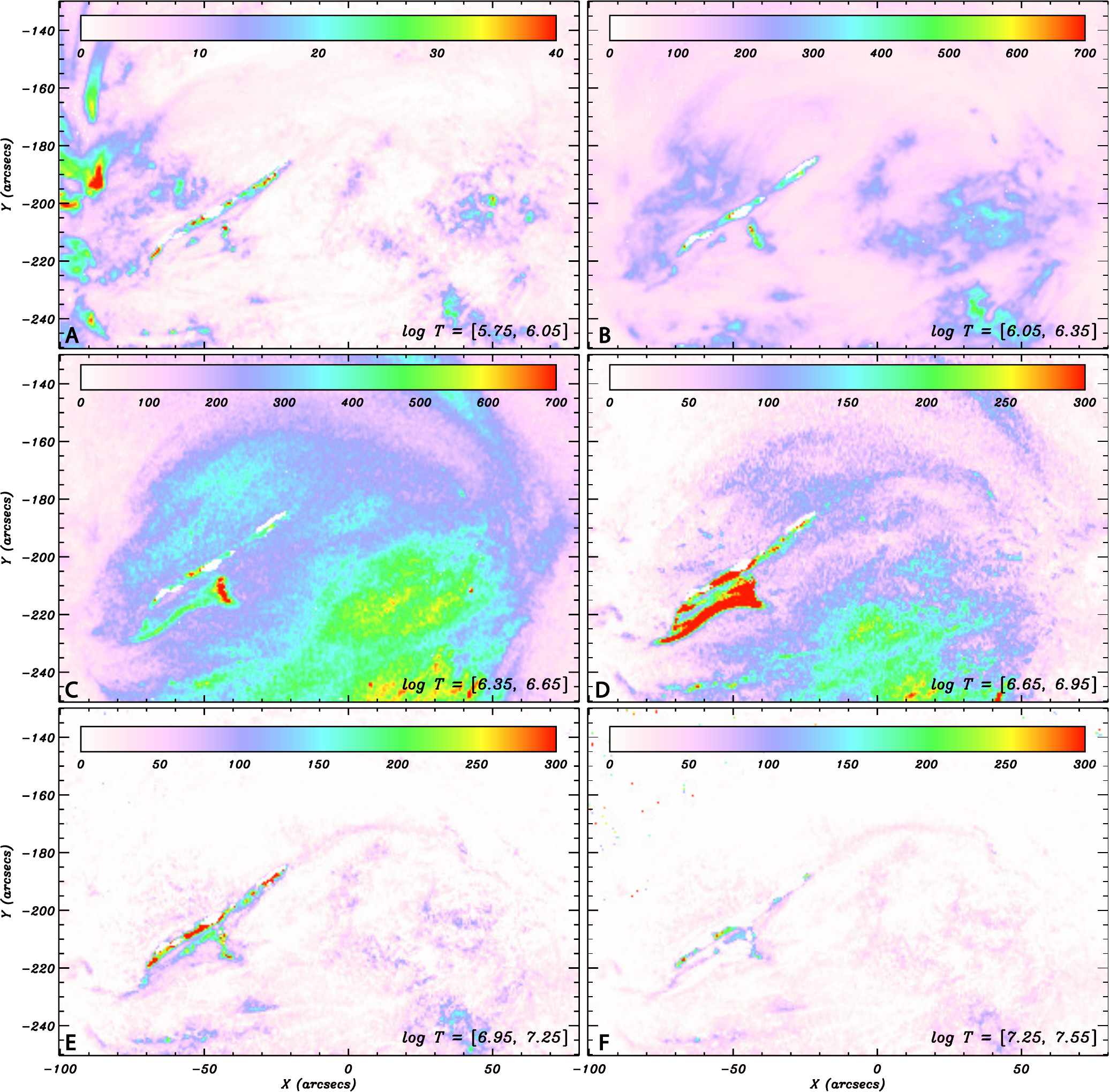}
  \caption{Emission measure maps at the peak time of evolution of the transient in different temperature bins as labelled. See also the corresponding animation. The unit for color bar in each panel are [$\times$10$^{26}$]~cm$^{-5}$. An animated version of this figure is available in the HTML version of the article, showing the evolution of EM in each temperature bin for 18 mins. The animation clearly shows the clumps in the transient structure and it multi-stranded nature. The highest emission measure is seen at the base in EM maps corresponding to the temperature of $\log\, T = 6.65- 6.95.$}. \label{em_maps}
\end{figure*}

The complete structure of the transient is best observed in the 94~{\AA} images of the AIA (see Figs.~\ref{hot} \& ~\ref{hot_94}). The transient comprises short loops located at the base and longer loops connecting the other main polarity of the active region. Transient events with a similar morphology have been reported earlier by \cite{Han_1997} and have been referred to as \enquote{double loop configuration of solar flares.} The AIA 304~{\AA} images also reveals the presence of a twisted filament (panel D of Fig~\ref{cool}). By combining the observations recorded in cool channels such as 304~{\AA} (see the corresponding animation Figure3D\_animation.mp4) with those in 193 or 94~{\AA} images, we find that the transient brightening appears in different strands. Observations also reveal that these strands are winding around the filament. Moreover, the transient event is highly filamentary (see Figs.~\ref{hot_94} and \ref{em_maps}), which is suggestive of multi-stranded structure. The highest emission is observed at the base of the transient within the temperature bin of $\log\, T=6.65 - 6.95$ (see panel D of Fig.~\ref{em_maps}), whereas the long loop-like structure is best seen, albeit very faint, in higher temperature $\log\, T=6.95 - 7.55$ (panels E \& F of Fig.~\ref{em_maps}). However, we note that such emission in high temperatures should be interpreted with caution due to limited constraints at higher temperatures \citep[e.g.][]{WinWS_2012}. Moreover, the structures observed in the EM maps obtained at such high temperatures may also be due to the presence of unaccounted cooler lines in the passbands of 94~{\AA} and 131~{\AA} passbands as was suggested by \cite{YouM_2014}.

Combining the AIA observations with those from HMI line-of-sight magnetograms, we find that the base of the transient is rooted in primary negative polarity and the parasitic positive polarity, whereas the long loop connects to the main positive polarity of the active region (see Fig.~\ref{hmi_aia}). The evolution study of the photospheric LOS magnetic field throughout the existence of the transient, i.e., for 19 mins, did not show any noticeable change in the structure of the magnetic field at the base of the transient (see Fig.~\ref{hmi_mag}). However, over a long time scale, we observed noticeable changes in the field configuration (see Figure9\_animation.mp4) as well as a change in the total signed flux over a small boxed region located at the base of the transient. Therefore, it is plausible to conclude that changes in the magnetic flux may have resulted in the transient. However, the data do now allow a conclusive determination of whether flux changes caused the observed transient.

\begin{figure*}
\centering
\includegraphics[width=0.3\textwidth]{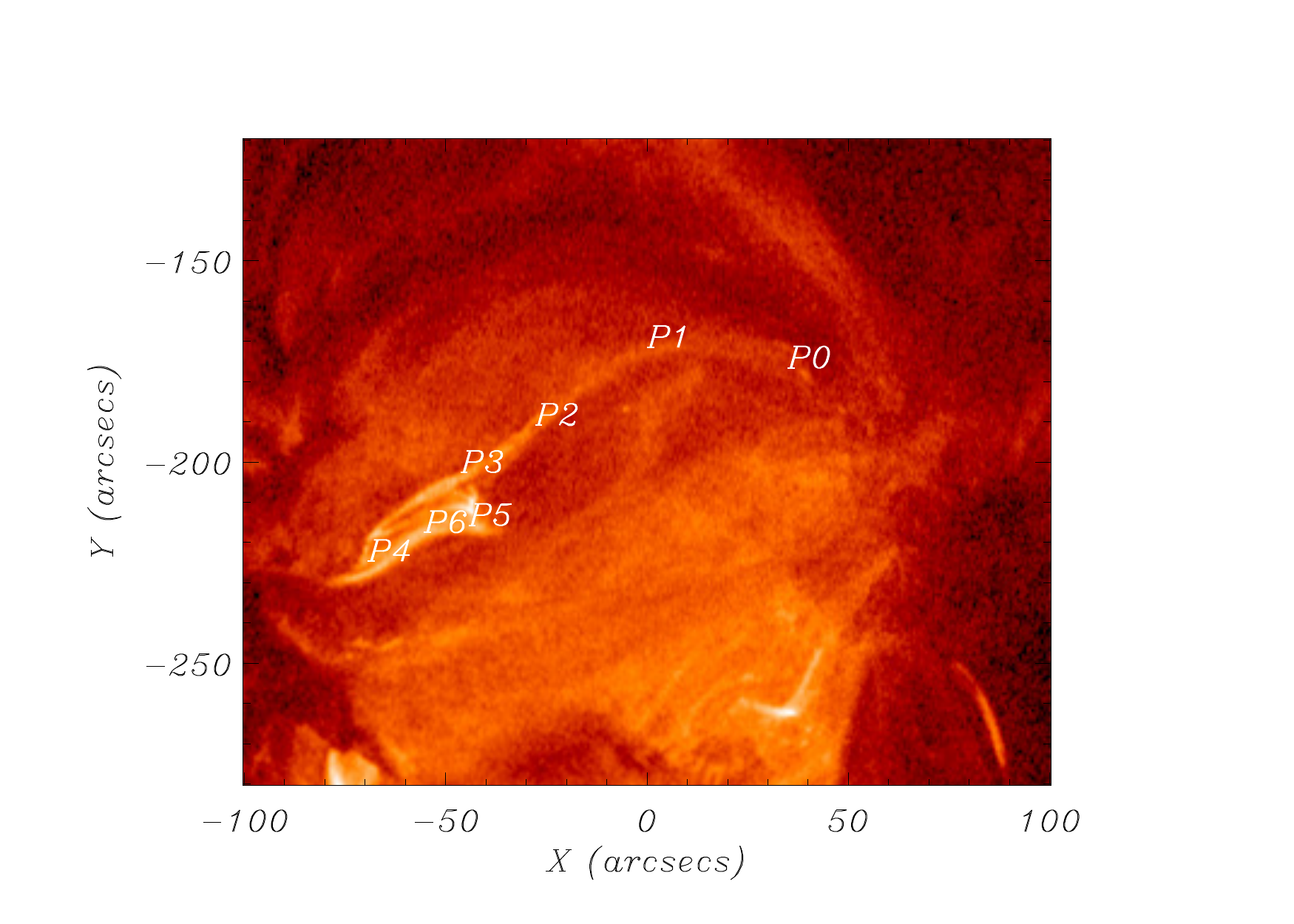}
\includegraphics[width=0.3\textwidth]{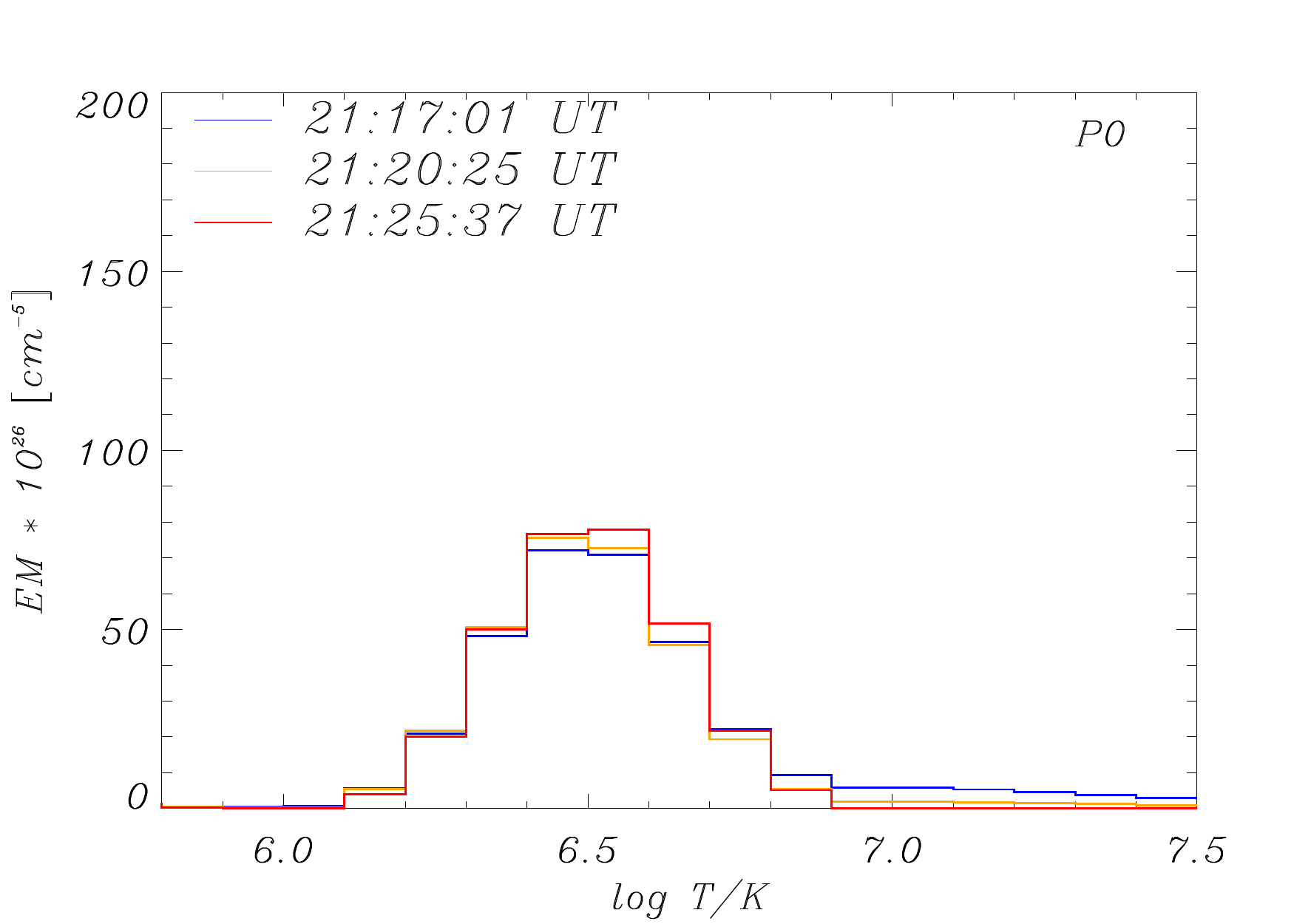}
\includegraphics[width=0.3\textwidth]{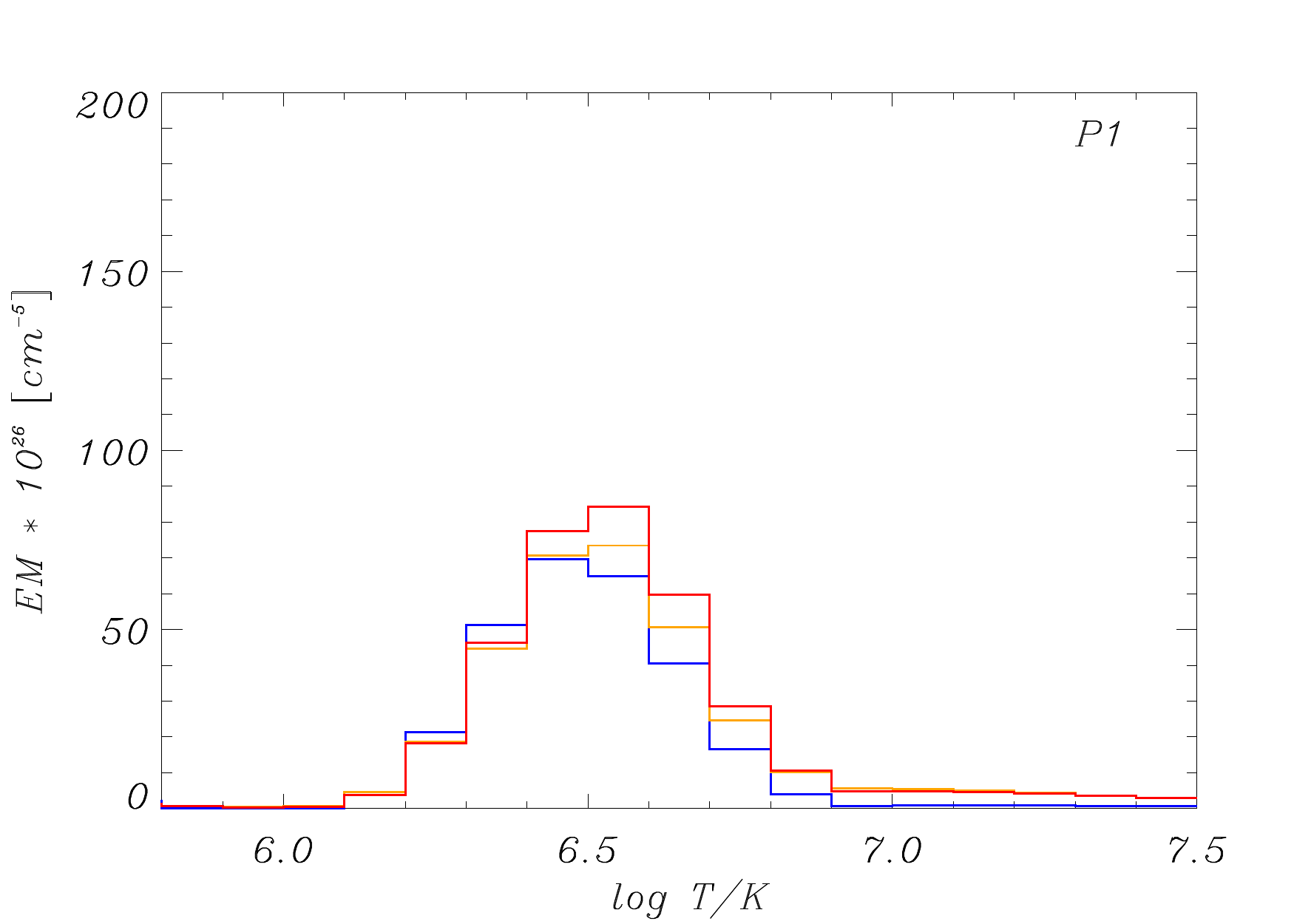}
\includegraphics[width=0.3\textwidth]{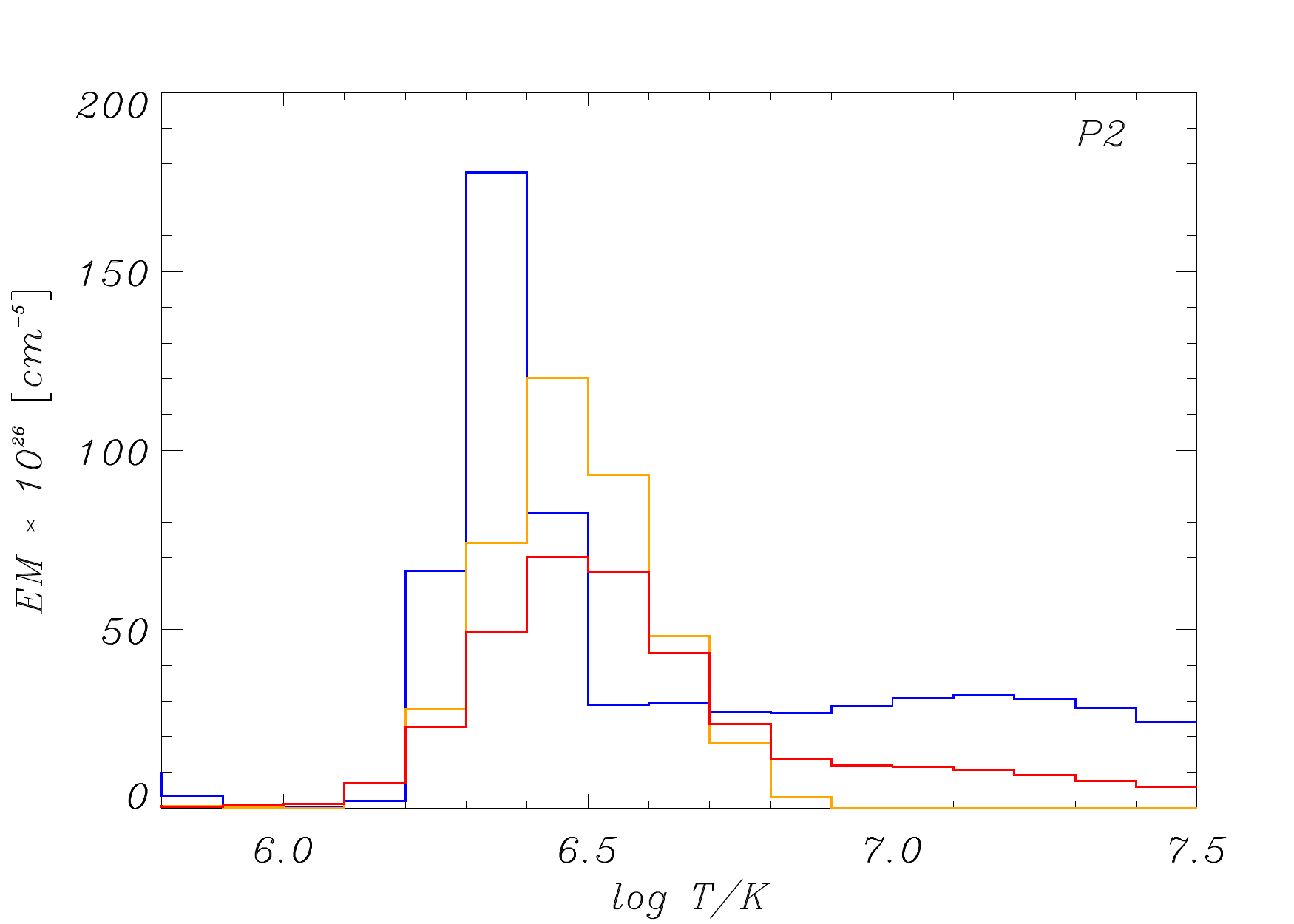}
\includegraphics[width=0.3\textwidth]{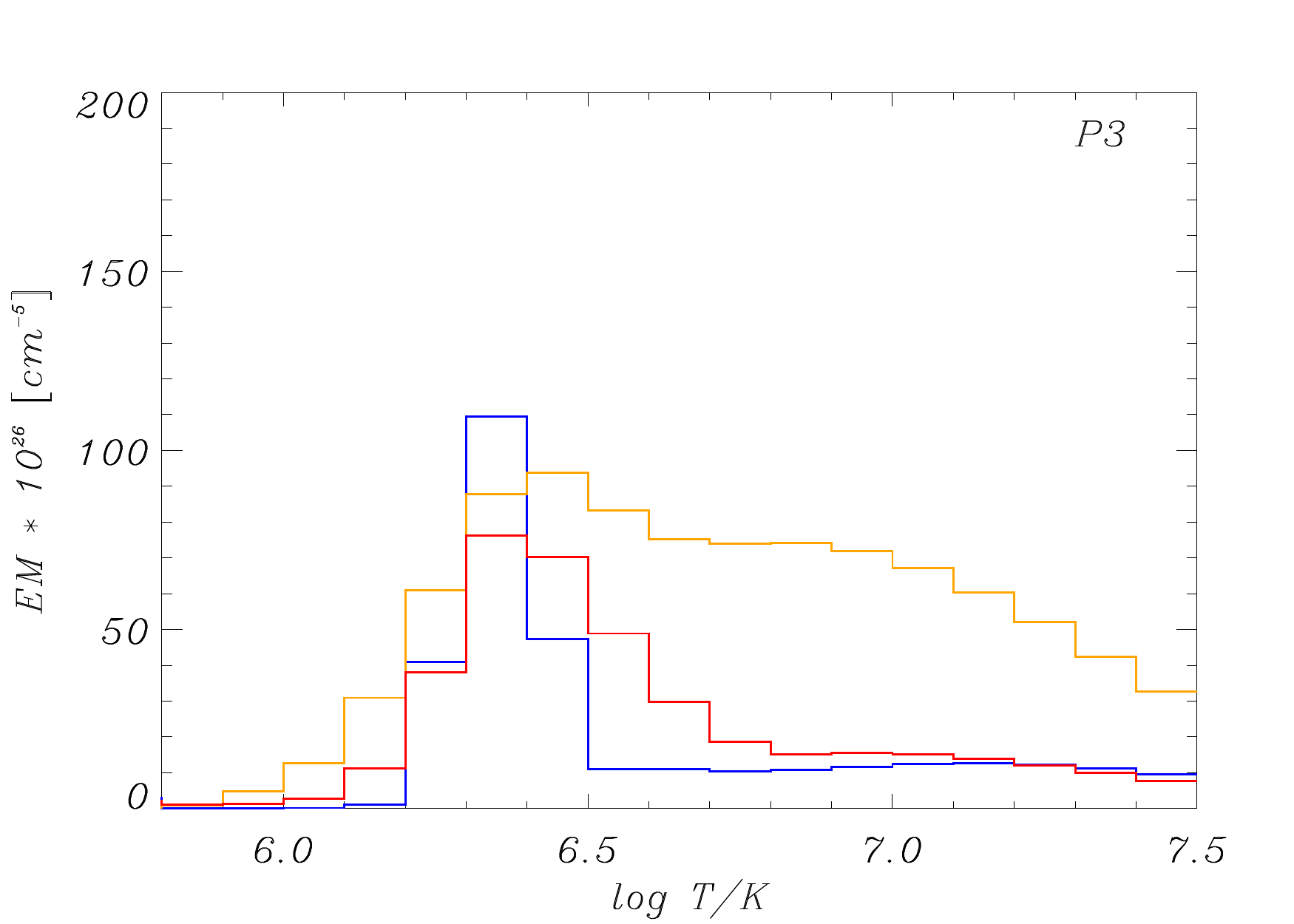}
\includegraphics[width=0.3\textwidth]{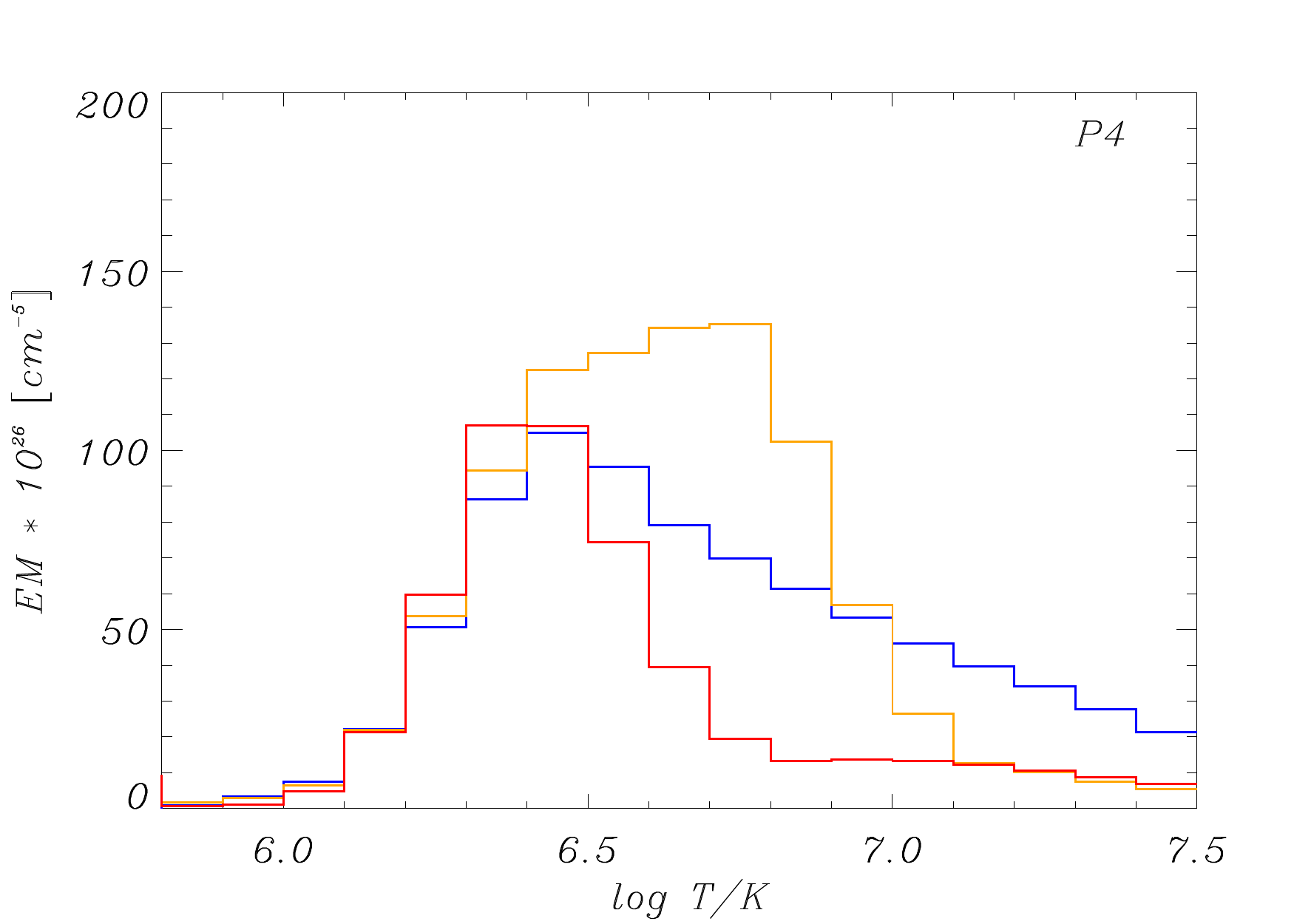}
\includegraphics[width=0.3\textwidth]{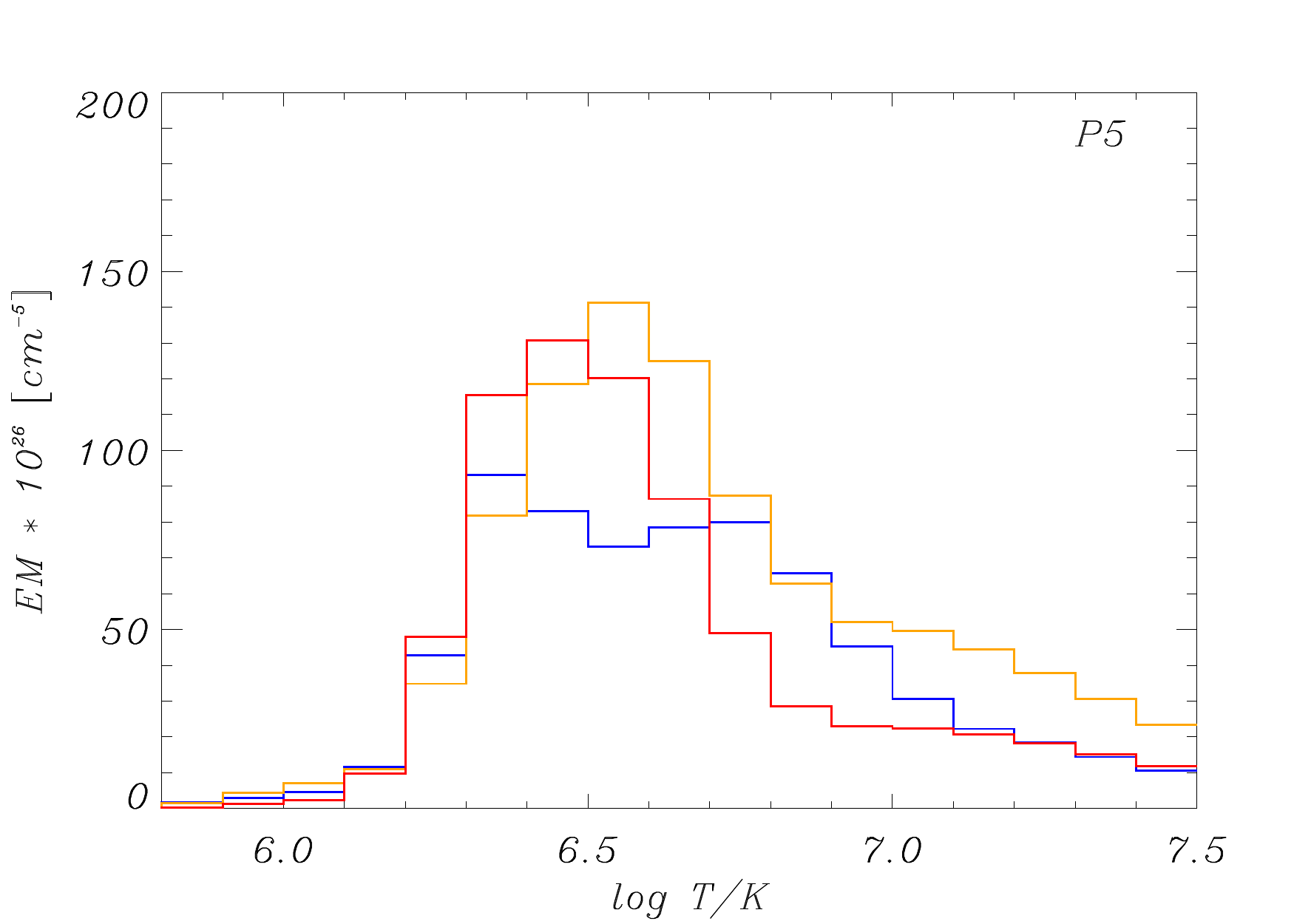}
\includegraphics[width=0.3\textwidth]{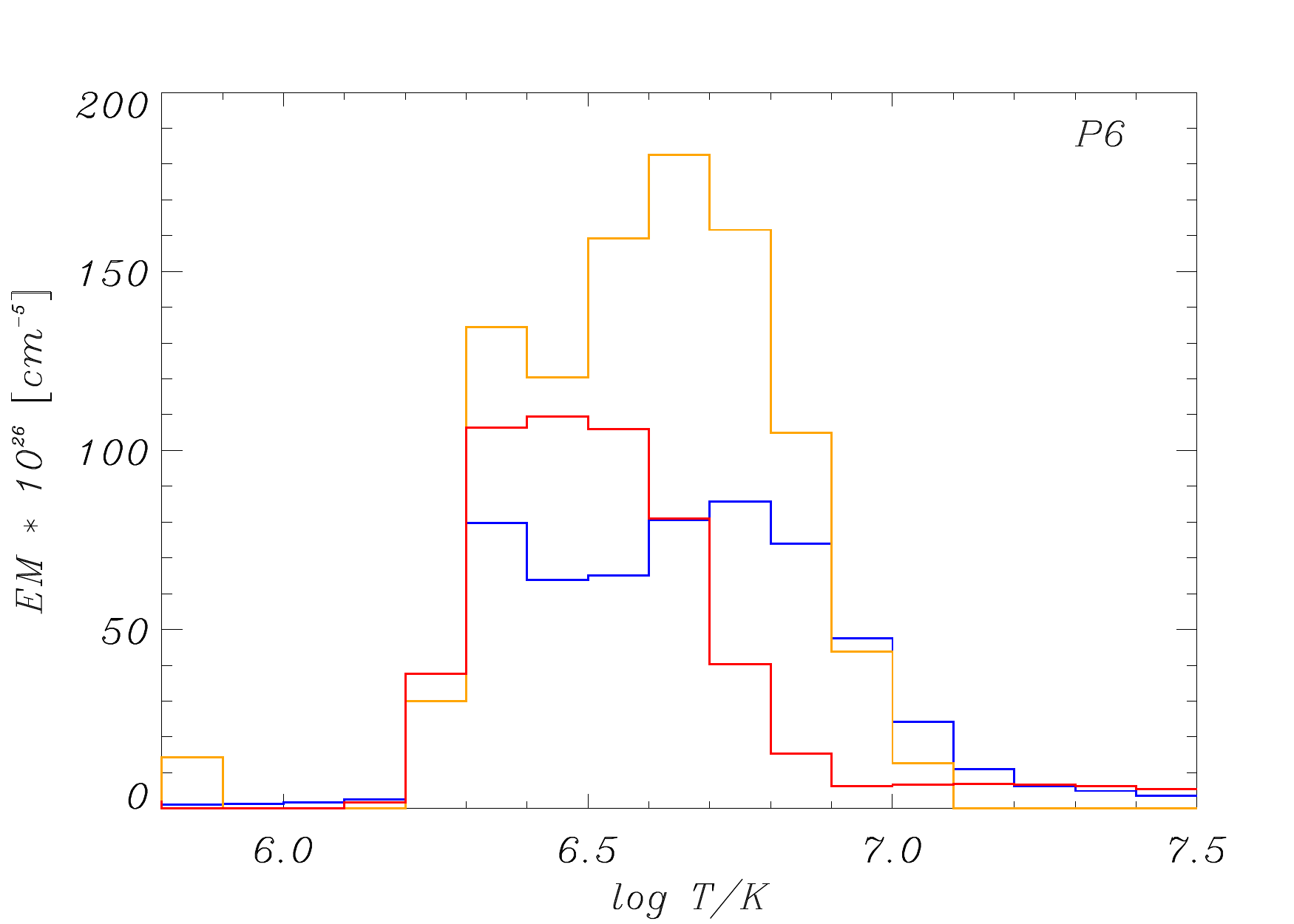}
\caption{Emission measure as a function of temperature at seven different locations along the transient obtained during the early phase (blue), peak phase (orange) and towards the end of the transient (red).}\label{em_plots}
\end{figure*}

\begin{figure}
\centering
\includegraphics[width=0.45\textwidth]{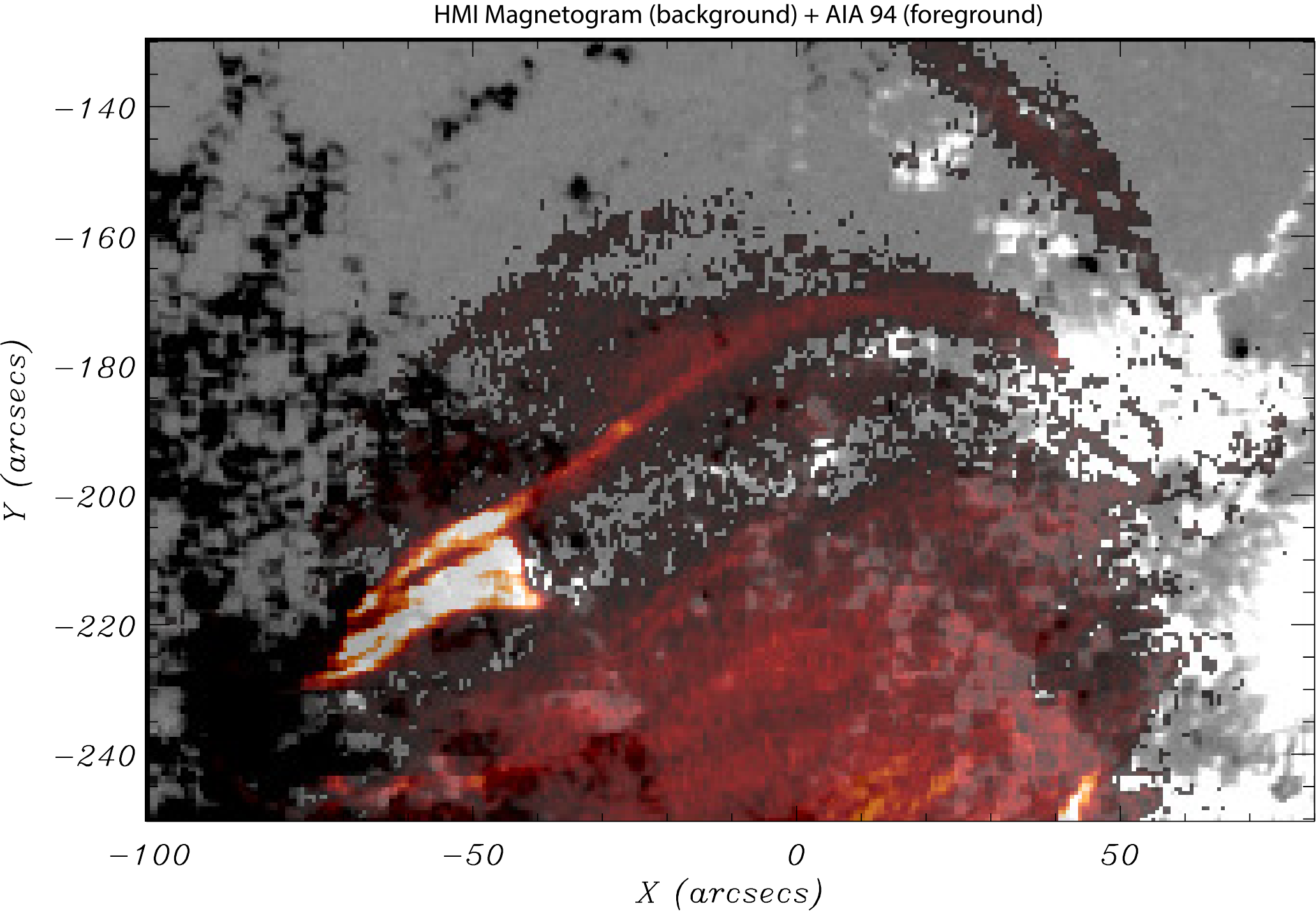}
\includegraphics[trim={0.7cm 1.1cm 1.2cm 1.3cm},clip, width=0.5\textwidth]{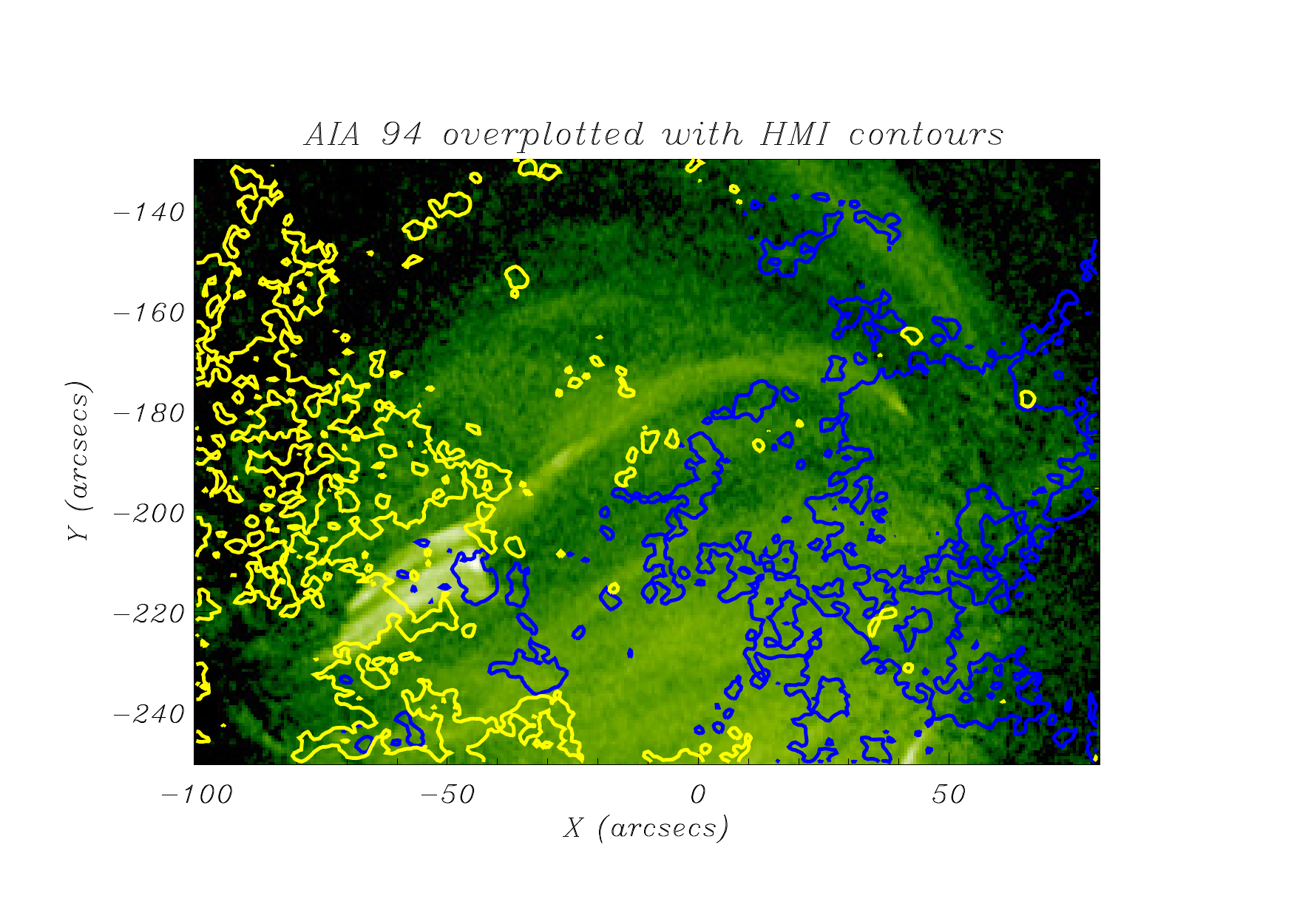}
\caption{AIA and HMI images taken nearly simultaneously. Left panel: \ion{Fe}{18} map blended with HMI LOS magnetograms. Right panel: \ion{Fe}{18} map over-plotted with magnetic field contours. Blue represents the positive magnetic polarity whereas the yellow represent the negative magnetic polarity. The contour HMI levels are $\pm$~100 G.} \label{hmi_aia}
\end{figure}

\begin{figure*}
\centering
\includegraphics[width=0.85\textwidth]{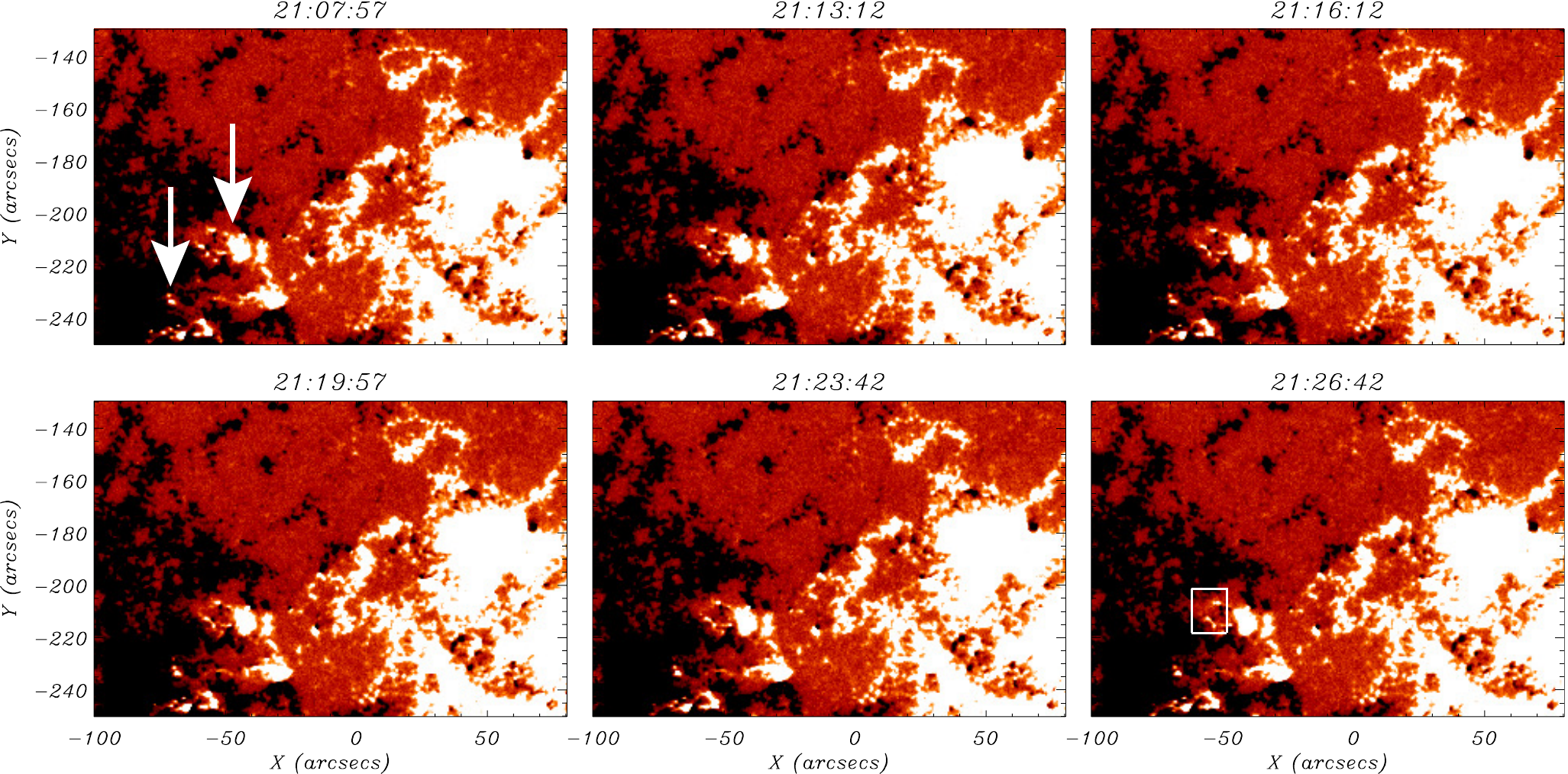}
\caption{Sequence of HMI LOS magnetograms showing the evolution of magnetic flux density in the source regions. The arrows locate the regions where the base of the transient is located. The box the bottom right panel shows the region which we considered for quantitative study of the evolution of magnetic flux with time. The magnetic field maps are displayed in the range of $\pm$100G. An animated version of this figure is available in the HTML version of the article, demonstrating the evolution of the LOS photospheric magnetic flux at the base of the transient, starting three hours before the transient's onset. Over a period of three hours, there is a substantial change in the photospheric magnetic flux at the base. Note that to enhance the small-scale flux, the maps are displayed between $\pm$40~G.} \label{hmi_mag}
\end{figure*}

\begin{figure*}
\centering
\includegraphics[width=0.85\textwidth]{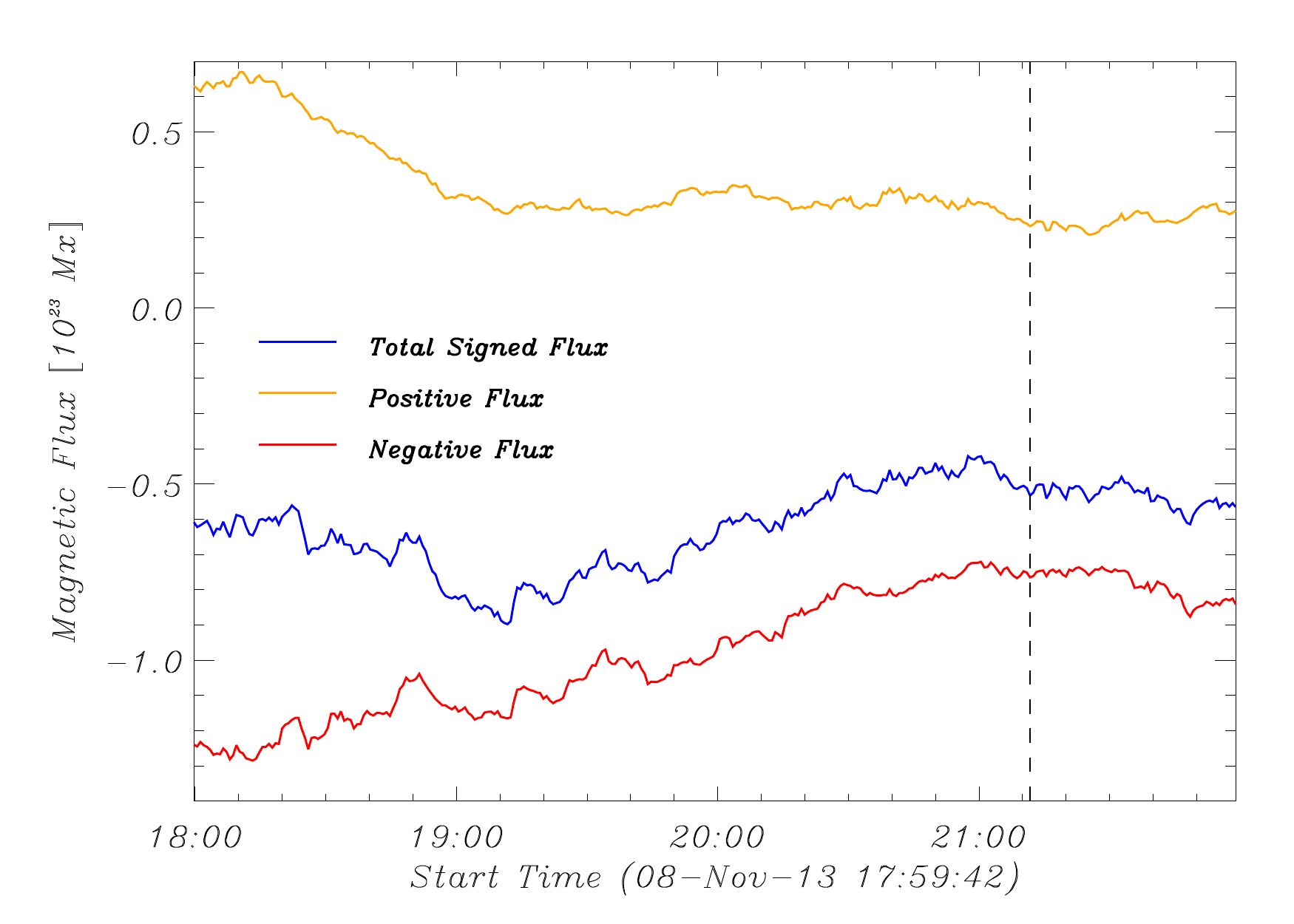}
\caption{Evolution of photospheric line-of-sight magnetic flux with time, computed over the box region shown in the bottom right panel of Fig.~\ref{hmi_mag}. Blue is for total signed flux whereas orange (red) is for positive (negative) flux.} \label{mag_flux}
\end{figure*}

\begin{figure*}
\centering
\includegraphics[trim=100 900 50 300,clip, width=0.85\textwidth]{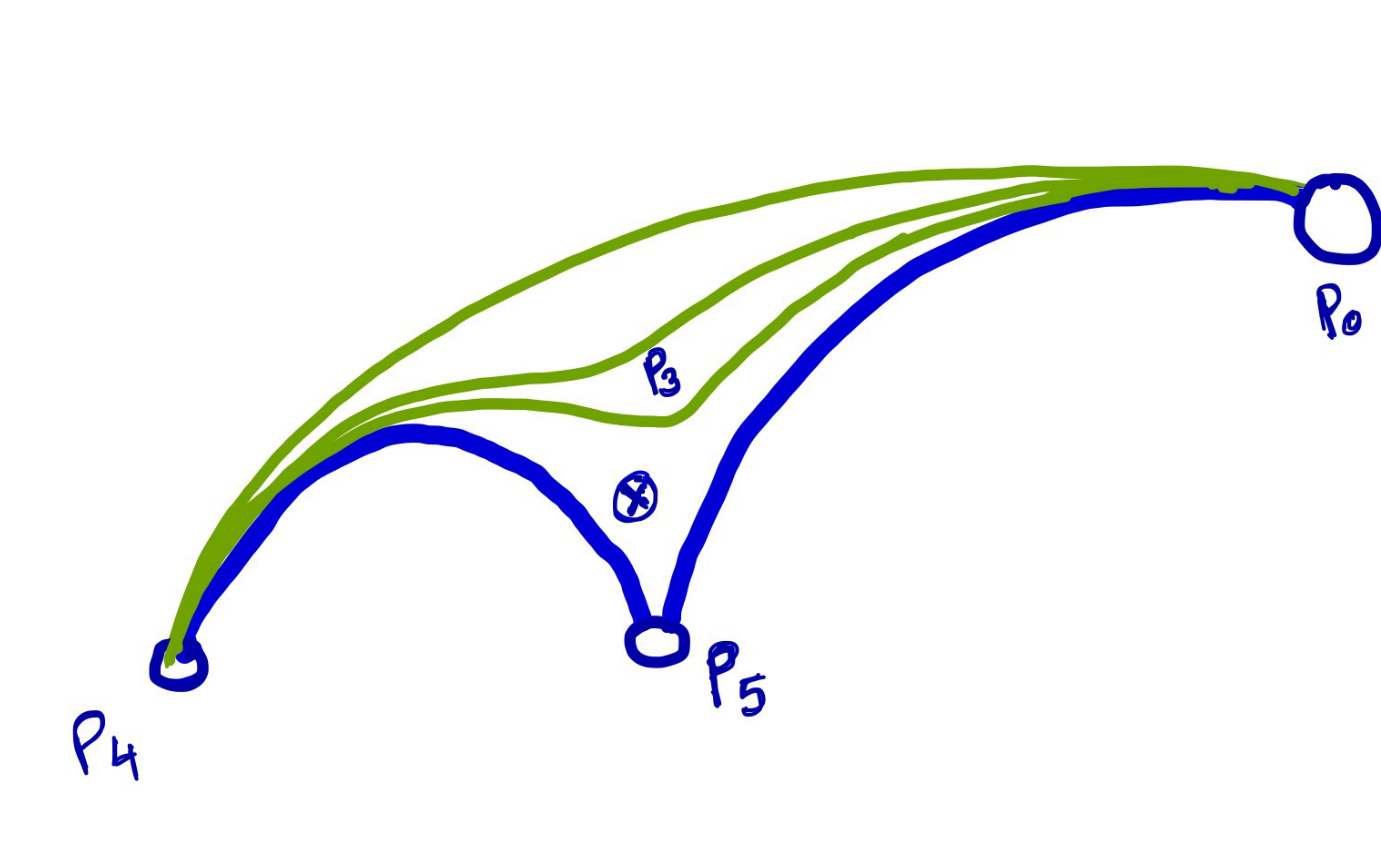}
\caption{Schematic representation of the transient. The points P0, P3, P4 and P5 represent the corresponding locations shown in the top left panel of Fig.~\ref{em_plots}. The encircled cross marks the possible location of reconnection between the loops connecting P0 {--} P5 and P4 {--} P5, shown in blue. The green lines represent the reconnected field lines. Location P3 depicts the kinked field lines, also observed in Fig.~\ref{hot}.} \label{cart}
\end{figure*}

Motivated by the observations, we provide a schematic diagram in Fig.~\ref{cart}, which is similar to that of \cite{Han_1997} and depicts the evolution of the transient studied here. We envision that there are small loops structures rooted within the main polarity forming between P4 and P5. These loops are due to the presence of parasitic polarities. There are also other loop systems connecting the two main polarities from P0 to P5. In Fig.~\ref{cart}, we show these loops in blue. Our observations suggest that reconnection occurs between the blue loops, leading to loops forming with green color. The loops coming out from the reconnection region will have a kink, which is also supported by the observational results of the presence of the kink in Fig.~\ref{hot} marked with an arrow. Moreover, the EM curves suggest a substantial enhancement in temperature at the kink location (see Fig.~\ref{em_plots}, P3). 

It is important to note that the physics of the formation of these transients may be similar to those of jets in different topological scenarios \citep[see][for a review on jets]{RaoPP_2016}. In the typical coronal jet scenario, the reconnection occurs among closed field lines and open field lines giving rise to a collimated structure. However, the observation reported here suggests reconnection between two closed field lines that leads to the formation of a longer loop, which is analogous to the spire of typical coronal jets, similar to those observed in the simulations of \cite{GonAT_2009, ArcTG_2010}. Moreover, the properties such as lifetime (15 min), multi-thermal nature, and DEM at the base of the transient observed here are similar to those obtained for jets observed at the periphery of active regions \citep[see, e.g.,][]{MulTDM_2016, RaoPP_2016}. We further note that the presence of filament at the base of the transients is also similar to jet events reported by many authors \citep[see, e.g.][and references therein]{HonJR_2011, SteRD_2015, HonJY_2016, PanSM_2017, KumKA_2018}.

Although we have qualitatively described the formation of the transient reported here, further observational work, primarily with spectroscopic measurements combined with modeling, is necessary for a complete understanding of the physics of such transients within the core of active regions. Such studies will help comprehend the formation of hot plasma in the core of active regions.

\acknowledgments
I thank the referee for critical comments that helped improve the paper. This research is partly supported by the Max-Planck Partner Group on the \textit{Coupling and Dynamics of the Solar Atmosphere} of MPS at IUCAA. IRIS is a NASA small explorer mission developed and operated by LMSAL with mission operations executed at NASA Ames Research centre and major contributions to downlink communications funded by the Norwegian Space Centre (NSC, Norway) through an ESA PRODEX contract. AIA and HMI data are courtesy of SDO (NASA). Facilities: SDO (AIA, HMI).
\bibliography{reference} 

\end{document}